\documentclass[%
 aip,
 amsmath,amssymb,
 reprint,%
]{revtex4-1}

\usepackage{graphicx}
\usepackage{dcolumn}
\usepackage{bm}

\usepackage[utf8]{inputenc}
\usepackage[T1]{fontenc}
\usepackage{mathptmx}
\usepackage{etoolbox}

\makeatletter
\def\@email#1#2{%
 \endgroup
 \patchcmd{\titleblock@produce}
  {\frontmatter@RRAPformat}
  {\frontmatter@RRAPformat{\produce@RRAP{*#1\href{mailto:#2}{#2}}}\frontmatter@RRAPformat}
  {}{}
}%

\newcommand{\fref}[1]{Fig.~\ref{#1}}

\newcommand{\sref}[1]{Sec.~\ref{#1}}

\newcommand{\dd}{\mathrm{d}}

\def\apj{Astrophys. J.}
\def\apjl{Astrophys. J. Lett.}

\def\prd{Phys. Rev. D}
\def\prl{Phys. Rev. Lett.}

\def\baas{Bull. Am. Astron. Soc.}
\def\nat{Nature}

\begin{document}

\preprint{AIP/123-QED}

\title[Black hole spectroscopy with ground-based atom interferometer and space-based laser interferometer gravitational wave detectors]{Black hole spectroscopy with ground-based atom interferometer and space-based laser interferometer gravitational wave detectors}

\author{Alejandro Torres-Orjuela}
 \email{atorreso@hku.hk}
 \affiliation{Department of Physics, The University of Hong Kong, Pokfulam Road, Hong Kong}

\date{\today}

\begin{abstract}
Gravitational wave (GW) detection has enabled us to test General Relativity in an entirely new regime. A prominent role in tests of General Relativity takes the detection of the Quasi-normal modes (QNMs) that arise as the highly distorted remnant formed after the merger emits GWs until it becomes a regular Kerr BH. According to the no-hair theorem, the frequencies and damping times of these QNMs are determined solely by the mass and spin of the remnant BH. Therefore, detecting the QNMs offers a unique way to probe the nature of the remnant BH and to test General Relativity. We study the detection of a merging binary black hole (BBH) in the intermediate mass range, where the inspiral-merger phase is detected by space-based laser interferometer detectors TianQin and LISA while the ringdown is detected by the ground-based atom interferometer (AI) observatory AION. The analysis of the ringdown is done using the regular broadband mode of AI detectors as well as using the resonant mode where the detection band is optimized to the frequencies of the QNMs predicted from the inspiral-merger phase. We find that using the regular broadband mode allows constraining the parameters of the BBH with relative errors of at most $10^{-6}$ from the ringdown while the frequencies and the damping times of the QNMs can be determined with total errors below $0.2\,{\rm Hz}$ and $115\,{\rm \mu s}$, respectively. Furthermore, we find that using the resonant mode can improve the parameter estimation for the BBH from the ringdown by up to one order of magnitude. Utilizing the resonant mode significantly limits the detection of the frequency of the QNMs but improves the detection error of the damping times by one to four orders of magnitude.  
\end{abstract}

\maketitle

\section{Introduction}\label{sec:int}

Numerous gravitational wave (GW) events have been detected during the first three observation rounds of the LIGO-Virgo-KAGRA Collaboration (O1-O3) and many more are expected from the ongoing round O4~\cite{GWTC1,GWTC2,GWTC3}. These detections had and have a profound impact on fundamental physics as they allow us to study gravity in the highly nonlinear and dynamical regime present during the merger of two black holes (BHs) and other compact objects. The data collected during these events enable us to test General Relativity, which has passed all previous experimental tests with flying colors, in an entirely new regime, complementing existing laboratory
and astrophysical tests of General Relativity~\cite{will_2014,berti_barausse_2015}. A possible way to examine the validity of General Relativity is ringdown tests that probe the consistency of the merger dynamics with the predictions of a Kerr BH as the remnant object~\cite{dreyer_kelly_2004,berti_cardoso_2006,gossan_veitch_2012,meidam_agathos_2014,ligo_virgo_2021e,ligo_virgo_2021f,zi_zhang_2021}. Particularly important for tests involving the ringdown phase is the detection of the Quasi-normal modes (QNMs) that arise as the highly distorted remnant formed after the merger emits GWs until it settles down to a regular Kerr BH~\cite{kokkotas_schmidt_1999}. According to the no-hair theorem of General Relativity, the QNM's frequency and damping time are determined by only the mass and spin of the remnant BH~\cite{penrose_1969,carter_1971,hansen_1974,gurlebeck_2015} and, hence, their detection offers a unique way to probe the nature of the remnant~\cite{dreyer_kelly_2004,berti_cardoso_2006}.

Existing ground-based laser interferometer observatories LIGO~\cite{ligo_2015}, Virgo~\cite{virgo_2012}, and KAGRA~\cite{kagra_2019} as well as ongoing detection campaigns with Pulsar Timing Arrays such as NANOGrav~\cite{arzoumanian_baker_2020}, the Chinese Pulsar Timing Array~\cite{cpta_2023}, the Parks Pulsar Timing Array~\cite{ppta_2023}, and the European Pulsar Timing Array~\cite{epta_2023} have been leading the path of GW astronomy. Current and future detectors are expected to continue expanding this field; opening new windows into the universe and allowing more stringent tests of General Relativity. Upcoming detectors include those based on laser interferometry  -- space-based observatories like LISA~\cite{lisa_2017}, TianQin~\cite{tianqin_2016}, Taiji~\cite{taiji_2015}, and DECIGO~\cite{decigo_2021} as well as future ground-based facilities such as the Einstein Telescope~\cite{et_2010} and Cosmic Explorer~\cite{cosmic_explorer_2019} -- and atom interferometer (AI) observatories~\cite{snadden_mcguirk_1998,dimopoulos_graham_2008,abend_allard_2023} such as AION~\cite{aion_2020}, ZAIGA~\cite{zaiga_2020}, and AEDGE~\cite{aedge_2020}.

AI detectors could take a prominent role in tests of General Relativity as they close the gap in the ${\rm dHz}$-band~\cite{abend_allard_2023} existing between space-based~\cite{tianqin_2016,tianqin_2021,lisa_2017,lisa_2024} and ground-based~\cite{ligo_2015,virgo_2012,kagra_2019} laser interferometer detectors, and thus allowing particularly accurate tests across multiple bands~\cite{carson_yagi_2020a,carson_yagi_2020b,jani_shoemaker_2020}. Such multi-band tests are particularly interesting as they allow coverage of longer fractions of the full inspiral-merger-ringdown signal involving very different regimes of gravity. Moreover, as different detectors come with different benefits and deficiencies, combining multiple detectors opens up the possibility of implementing optimized tests of General Relativity.

In this paper, we study the detection of a merging binary black hole (BBH) in the intermediate mass range. We analyze the detection of the inspiral-merger phase by space-based laser interferometer GW detectors TianQin~\cite{tianqin_2021} and LISA~\cite{lisa_2024} and of the QNMs using the noise curve of the AI detector AION-1km\cite{aion_2020}. In particular, we explore how well the starting time of the late phase (linear regime) of the ringdown and the frequencies of the QNMs can be predicted from the parameter estimation of the inspiral and merger. The analysis of these predictions is done, as we also study how the detection of the QNMs can improve when using the resonant mode of AI detectors tuned to the frequencies predicted from the inspiral-merger phase. The detection of the inspiral and/or the merger phase and the ringdown phase of merging BBHs by multiple detectors has been explored before~\cite{carson_yagi_2020a,carson_yagi_2020b,jani_shoemaker_2020}. However, to the best of our knowledge, this is the first work to make such a study involving AI detectors in the ${\rm dHz}$-band and, in particular, utilizing its resonant mode. Moreover, due to the proximity of this band with the ${\rm mHz}$-band of space-based laser interferometer detectors, the results obtained are particularly promising.

\section{Quasi-normal modes}\label{sec:qnm}

QNMs arise, as perturbations of BHs that lead to the emission of GWs~\cite{kokkotas_schmidt_1999}. In this emission, there are no normal mode oscillations but instead, the frequencies are ``quasi-normal'': complex-valued frequencies with the real part representing the actual frequency of the oscillation and the imaginary part representing the damping time of the emission. The natural way to study these oscillations is by considering perturbations of the linearized Einstein equations. The study of BH perturbations was initiated by Regge and Wheeler~\cite{regge_wheeler_1957} in the 1950s and continued by Zerilli~\cite{zerilli_1970} and many others later on. QNMs were first pointed out by Vishveshwara~\cite{vishveshwara_1970b} in calculations of the scattering of gravitational waves by a Schwarzschild black hole, while Press~\cite{press_1971} coined the term quasi-normal frequencies. Price, Pullin, and collaborators~\cite{price_pullin_1994,anninos_price_1995,gleiser_nicasio_1996,andrade_price_1997} were the first to show the agreement between full nonlinear numerical results and results from perturbation theory for the coalescence of two BHs thus proving the of power QNMs in the study of BBH mergers.

The highly distorted BH remnant formed from a BBH merger emits GWs until it settles down to a regular BH. This phase of the emission is referred to as the ringdown where its late stages can be expressed as a superposition of QNMs~\cite{vishveshwara_1970a,cunningham_price_1978}. According to General Relativity, for astrophysical BHs, the frequencies and damping times of the QNMs are determined by the mass and spin of the remnant BH~\cite{penrose_1969,carter_1971,hansen_1974,gurlebeck_2015}. Therefore, detecting the frequencies and damping times of the QNMs offers a unique test of the BH nature of the merger remnant~\cite{dreyer_kelly_2004,berti_cardoso_2006}. An electric charge of the remnant BH can theoretically also affect its properties but it is not expected to leave a detectable imprint on the ringdown~\cite{carullo_laghi_2022}.

The complex GW amplitude of the late stages of the ringdown can be expressed as a superposition of damped sinusoids~\cite{london_2018,ligo_virgo_2021f}
\begin{widetext}
\begin{equation}\label{eq:qnm}
    h_+(t) - ih_\times(t) = \frac{GM}{c^2D_L}\sum_{\ell=2}^\infty\sum_{m=-\ell}^\ell\sum_{n=0}^\infty A_{\ell mn}\exp\left(-\frac{t-t_0}{(1+z)\tau_{\ell mn}}\right)\exp\left(\frac{i\omega_{\ell mn}(t-t_0)}{1+z}\right)\,_{-2}S_{\ell mn}(\theta,\phi),
\end{equation}
\end{widetext}
where $(\ell,m,n)$ are the indices of the QNMs, $G$ is the gravitational constant, $c$ is the speed of light in vacuum, $M$ is the total mass of the BBH in the observer frame, $t_0$ is the starting time of the ringdown, $z$ is the cosmological redshift of the source, and $D_L$ is the sources luminosity distance. $\omega_{\ell mn}$, $\tau_{\ell mn}$, and $A_{\ell mn}$ are the frequency, the damping time, and the excitation amplitude of the QNMs, respectively. We write $_{-2}S_{\ell mn}(\theta,\phi)={_{-2}S_{\ell m}^{\hat{a}_f\omega_{\ell mn}}(\theta,\phi)}$ where $_{-2}S_{\ell m}^{\gamma}(\theta,\phi)$ are spin-weighted spheroidal harmonics that encode the sky localization of the observer $(\theta,\phi)$ in spherical coordinates for a frame centered on the remnant BH and aligned with its angular momentum, and $\hat{a}_f$ is the dimensionless spin magnitude of the remnant BH~\cite{berti_cardoso_2006}. The index $n$ denotes various tones of the spectrum starting with the basic/strongest tone $n=0$.

We use the $A_{\ell mn}$ from London (2018)~\cite{london_2018} and the $\omega_{\ell mn}$, $\tau_{\ell mn}$, and spherical-spheroidal mixing coefficients from London \& Fauchon-Jones (2019)~\cite{london_fauchon-jones_2019}. Note that the model introduced in London (2018)~\cite{london_2018} only applies to non-precessing BBHs. Therefore, we use the Numerical Relativity surrogate remnant model \texttt{NRSur3dq8Remnant} which can predict the remnant mass $M_f$ and the remnant spin $\hat{a}_f$ for non-precessing BBHs with mass ratios $q\leq8$ and spin magnitudes $|\hat{a}_1|,|\hat{a}_2|\leq0.8$~\cite{varma_gerosa_2019}. Moreover, we generate the spin-weighted spheroidal harmonics using the Python package \texttt{spheroidal}~\cite{spheroidal_2023}.

\begin{figure*}
\includegraphics[width=0.98\textwidth]{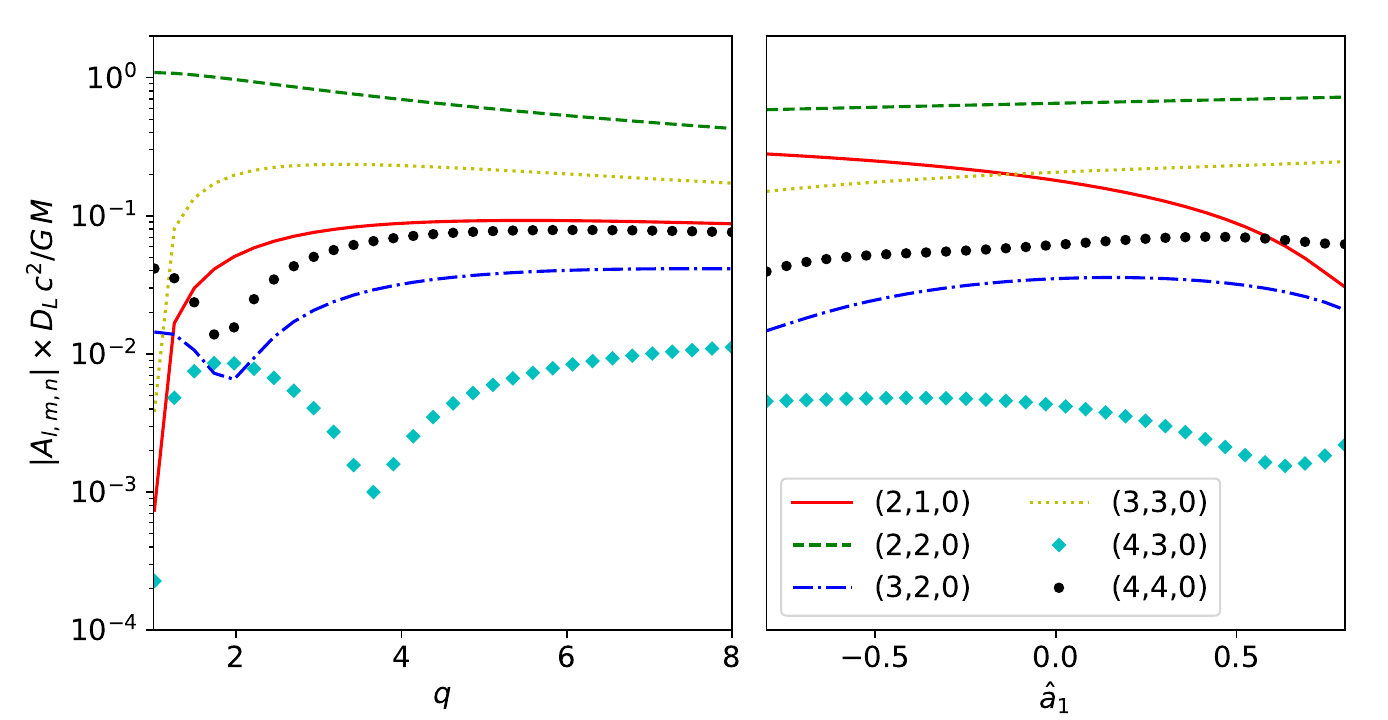}
    \caption{
        The absolute value of the amplitude $A_{\ell,m,n}$ of the six strongest QNMs. On the left-hand side, it is shown how the amplitudes vary as a function of the mass ratio $q$ while the right-hand side shows the amplitude as a function of the primary black hole's spin $\hat{a}_1$.
    }\label{fig:amp}
\end{figure*}

\begin{figure*}
\includegraphics[width=0.98\textwidth]{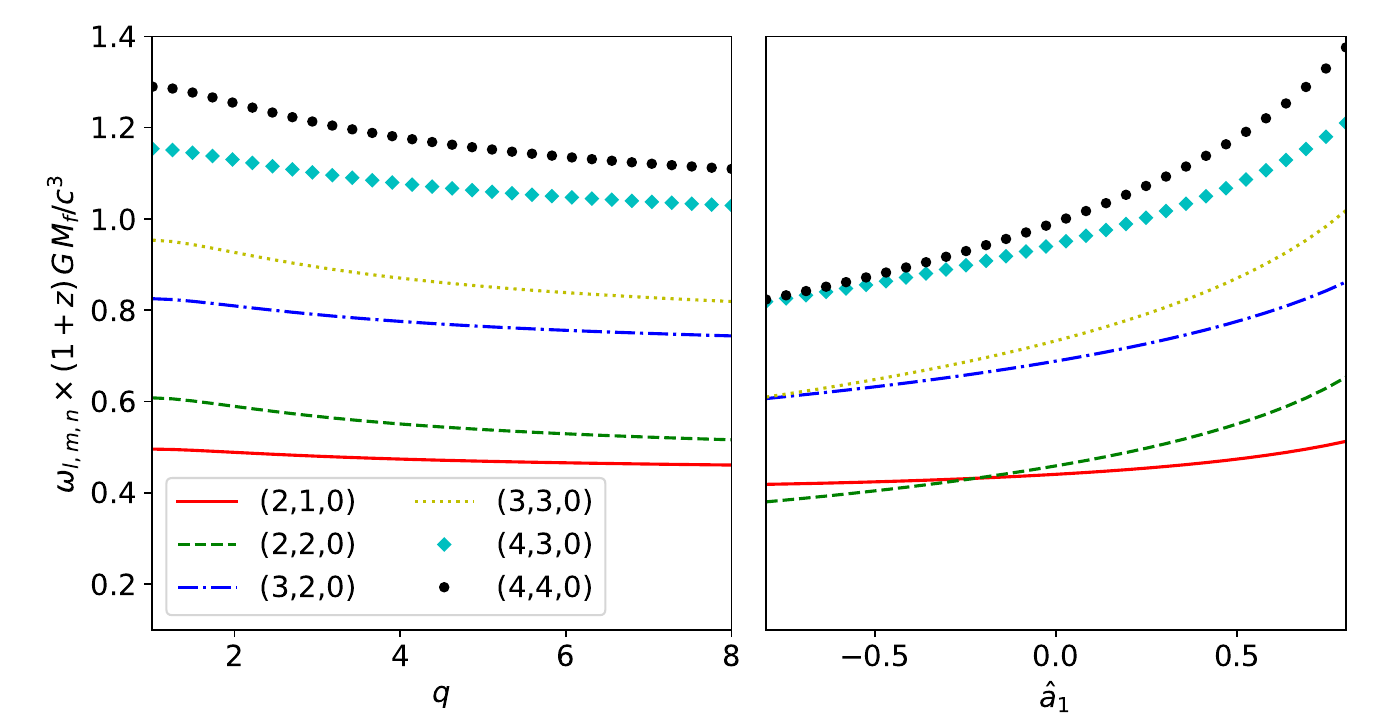}
    \caption{
        The frequency $\omega_{\ell,m,n}$ of the six strongest QNMs. The left-hand side shows the frequency as a function of the mass ratio $q$. The right-hand side depicts the frequency as a function of the primary black hole's spin $\hat{a}_1$.
    }\label{fig:fre}
\end{figure*}

\fref{fig:amp} shows the absolute value of the amplitude of the QNMs $A_{\ell,m,n}$ as a function of the BBH's mass ratio $q$ (left) and of the magnitude of the spin of the primary BH $\hat{a}_1$ (right). When varying the mass ratio, we set $\hat{a}_1=\hat{a}_2=0.5$ while we set $q=4$ and $\hat{a}_2=0.5$ when varying the spin. We see that in both cases $(2,2,0)$ is the strongest QNM and that for most parameters considered $(2,1,0)$ and $(3,3,0)$ are the second strongest modes. However, in some cases, the sub-dominant modes can reach amplitudes that are comparable to the amplitude of the stronger modes. The frequencies of the QNMs are shown in \fref{fig:fre}. We see that the QNMs with the same $\ell$ have relatively similar frequencies while there is a significant difference for varying $\ell$, and the QNMs $(2,1,0)$ and $(2,2,0)$ have the lowest frequencies of all modes. As we discuss in \sref{sec:mbbhs} in more detail, it is difficult to detect the six strongest QNMs while having a significant using the resonant mode of AI observatories. Therefore, we focus on the detection of the $(2,1,0)$ and $(2,2,0)$ modes in our analysis.

Note that London (2018)~\cite{london_2018} sets $t_0 =  20\,MG/c^3$ for the initial time of the linear regime of the ringdown which we adopt in our analysis of the QNMs. Furthermore, as their model only applies to non-precessing BBHs we use the Numerical Relativity surrogate model \texttt{NRHybSur3dq8} which can generate waveforms for non-precessing BBHs with mass ratios $q\leq8$ and spin magnitudes $\hat{a}_1,\hat{a}_2\leq0.8$ for the inspiral-merger phase~\cite{varma_field_2019b}. We point out that in London (2018)~\cite{london_2018} $t=0$ corresponds to the time of the peak of the $(2,2)$ spherical mode while in \texttt{NRHybSur3dq8} $t=0$ is set to the time of the peak of the full wave. Therefore, $t=0$ is in principle different in the two models but because the $(2,2)$-mode is the dominant mode during the merger of the BBH the two times only differ marginally and we treat them as equal in our analysis.

\section{Gravitational wave detection with atom interferometer observatories and space-based laser interferometer detectors}\label{sec:det}

A way to understand the effect of GWs is that they induce a change of a `phase' proportional to the amplitude of the wave and depending on the orientation of the wave and the system affected~\cite{maggiore_2008}. This `phase' can be, e.g., the phase of a photon or of the internal state of an atom. Detectors using light are usually denoted as laser interferometer detectors while those using atoms are often denoted as atom interferometer (AI) detectors~\cite{weiss_2022,dimopoulos_graham_2008}. In this section, we introduce briefly the basic concepts of AI detectors, then we discuss how GW signals can be extracted from the data of AI or laser interferometer detectors and how a Fisher matrix analysis can be used to estimate the properties of the source.

GWs can be detected through differential phase measurement in multiple AIs formed by cold atoms in free-fall that are operated simultaneously using a common laser source~\cite{snadden_mcguirk_1998,dimopoulos_graham_2008,graham_hogan_2013,graham_hogan_2017,aion_2020}. The atom sources for the AIs are positioned along the length of a vertical vacuum system where laser pulses are used to drive transitions between the ground and excited states of the atoms, while also acting as beam splitters and mirrors for the atomic de Broglie waves to perform interferometry with them. A quantum superposition of a ground state and a clock state is generated in each AI. The phase imprinted along the two different paths depends on the phase of the laser pulses (de-)exciting the atoms and on the phase accumulated by the atoms themselves due to energy shifts. Therefore, the GW strain can be read after recombining the two paths and comparing the accumulated phase of the spatially separated AIs.

Besides the regular or broadband mode described above, AI detectors can also be operated in the so-called resonant mode which is accomplished using a series of $Q$ $\pi$-pulses where each drives a transition between the ground and the excited state~\cite{le-bellac_2006,graham_hogan_2016}. The resonant mode has the advantage that we can achieve a $Q$-fold enhancement of the detector's sensitivity although with the drawback that the detector's band reduces to a width of $\sim f_r/Q$ around the resonance frequency of the detector $f_r := \pi/T$ ($2T$: the interrogation time of the AI).

AI observatories cover the so-called intermediate band in the ${\rm dHz}$-regime~\cite{aion_2020,zaiga_2020,aedge_2020}. Therefore, they are ideal for detecting the QNMs of BBH with masses of the order $10^4-10^5\,{\rm M_\odot}$ (see \fref{fig:fdet}). However, sources in this mass range have their inspiral and merger phase in the ${\rm mHz}$-regime which is covered by space-based laser interferometer detectors such as TianQin or LISA~\cite{tianqin_2016,lisa_2024} (see \fref{fig:stra}). Therefore, in our analysis, we consider the case where the BBHs are detected by TianQin and LISA before the merger while the QNMs are detected using AI detectors.

AI and laser interferometer GW detectors use different techniques to `collect' the GW data. Nevertheless, the data they obtain can be analyzed using the same techniques such as match filtering. The only major difference is that AI detectors are sensitive to GWs along one spatial direction while laser interferometer GW detectors compare the effect of the GW along two non-parallel directions. Therefore, we can define the usual noise-weighted inner product~\cite{finn_1992}
\begin{equation}\label{eq:innprod}
    \langle h, h'\rangle := 2\Re\left[\int_0^\infty\frac{\tilde{h}(f)\tilde{h'}(f)}{S_n(f)}\dd f\right]
\end{equation}
where $S_n$ is the one-sided power spectral density (PSD) of the detector, and $\tilde{h}$ and $\tilde{h'}$ are the Fourier transforms of two time-domain waveforms $h, h'$, respectively. Note that the ringdown signal cannot be Fourier transformed using common techniques because the damping time is similar to the time of one period. Therefore, we Fourier transform the ringdown signal following the procedure in Berti et al. (2006)~\cite{berti_cardoso_2006}.

For a signal $h$, the optimal signal-to-noise ratio (SNR) is obtained by filtering with the same waveform~\cite{finn_1992}
\begin{equation}\label{eq:snr}
    \rho^2 := \langle h, h\rangle.
\end{equation}
In the high SNR limit, linearized estimates for the measurement errors of the source's parameters $\mathbf{\theta}$ can be obtained using a Fisher matrix analysis~\cite{coe_2009}. For a waveform $h(\mathbf{\theta})$, the Fisher matrix is defined as
\begin{equation}\label{eq:deffm}
    \Gamma_{i,j} := \left\langle\frac{\partial h(\mathbf{\theta})}{\partial\theta_i},\frac{\partial h(\mathbf{\theta})}{\partial\theta_j}\right\rangle.
\end{equation}
By inverting the Fisher matrix $C := \Gamma^{-1}$ we obtain an approximate of the sample covariance matrix of the Bayesian posterior distribution. Therefore, the detection accuracy of the parameter $\mathbf{\theta}_i$ at the $1-\sigma$ level is given by the square root of the diagonal element $\sqrt{C_{ii}}$ while the non-diagonal elements $C_{ij}$ indicate the correlation between the parameters $\mathbf{\theta}_i$ and $\mathbf{\theta}_j$~\cite{snecdecor_cochran_1991}.

\section{Multi-band detection and black hole spectroscopy}\label{sec:mbbhs}

The frequency and damping times of the QNMs emitted at the late ringdown phase of a merging BBH are determined by the mass and spin of the remnant BH~\cite{penrose_1969,carter_1971,hansen_1974,gurlebeck_2015}. The specific properties of the QNM depend on the mode considered while the spin of the remnant BH governs the relationship between the frequency and the damping time, and the mass of the remnant sets their overall scale. At the same time, using General Relativity and assuming the BBH merges in a vacuum, it is possible to predict the properties of the remnant BH based on the properties of the BBH~\cite{israel_1982,heusler_1996,carter_1971,varma_gerosa_2019}. Therefore, it is possible to ``predict'' the QNMs by measuring the parameters of the BBH before and/or during the merger. In this section, we study the detection of the QNMs by AI detectors and the detection of the inspiral-merger phase by space-based laser interferometer detectors to compare the parameter estimation we can obtain from the different detectors and to understand how well the QNMs can be predicted.

\begin{figure}
\includegraphics[width=0.48\textwidth]{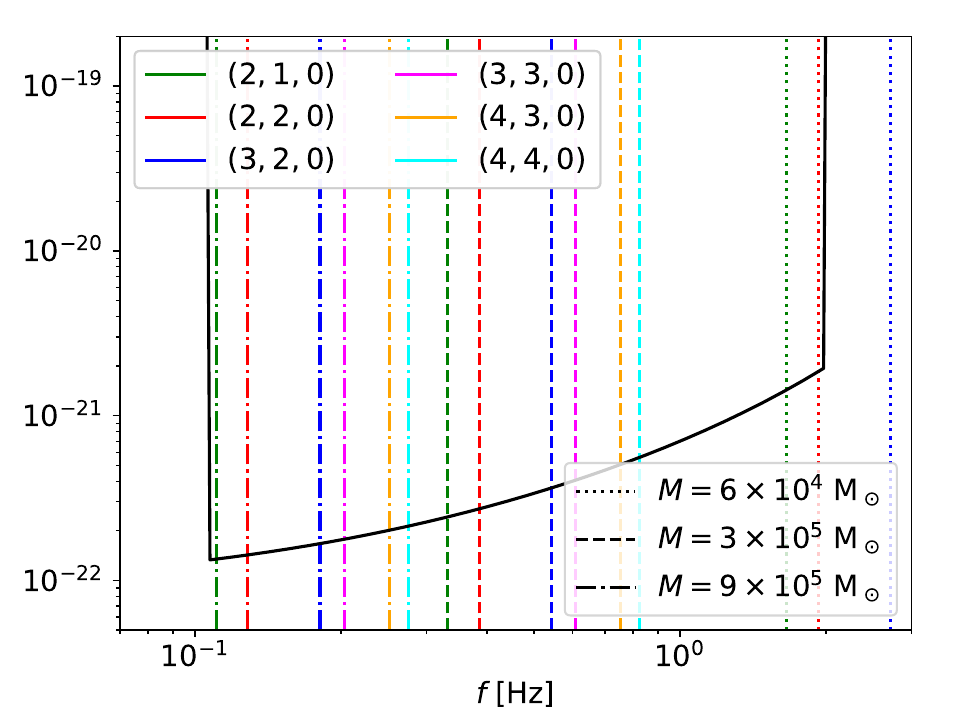}
    \caption{
        The frequencies of the six strongest QNMs for BBHs of different masses. The dotted lines show the frequencies for a BBH with $M=6\times10^4\,{\rm M_\odot}$, the dashed lines the frequencies for a BBH with $M=3\times10^5\,{\rm M_\odot}$, and the dashed-dotted lines the frequencies for a BBH with $M=9\times10^5\,{\rm M_\odot}$. Different colors represent the different QNMs and the black solid line shows the noise curve of AION.
    }\label{fig:fdet}
\end{figure}

In \fref{fig:fdet}, we show the frequencies of the six strongest QNMs for different masses compared to the noise curve, $\sqrt{S_n(f)}$, of AION-1km or short AION assuming the gravitational gradient noise can be fully modeled and mitigated~\cite{moore_cole_2015,aion_2020}. We see that the frequencies of all six modes cover around one-third of AION's band. Using the resonant mode of AI detectors we can greatly increase the sensitivity of the detector for certain frequencies at the cost of reducing the size of the band. This also means that the resonant mode is the most effective when detecting a small range of frequencies and, hence, we focus in our analysis only on the two dominant QNMs $(2,2,0)$ and $(2,1,0)$. We further see that to detect these two modes with AION the total mass of the BBH in the observer frame should be roughly in the range between $6\times10^4\,{\rm M_\odot}$ and $9\times10^5\,{\rm M_\odot}$.

\begin{figure}
\includegraphics[width=0.48\textwidth]{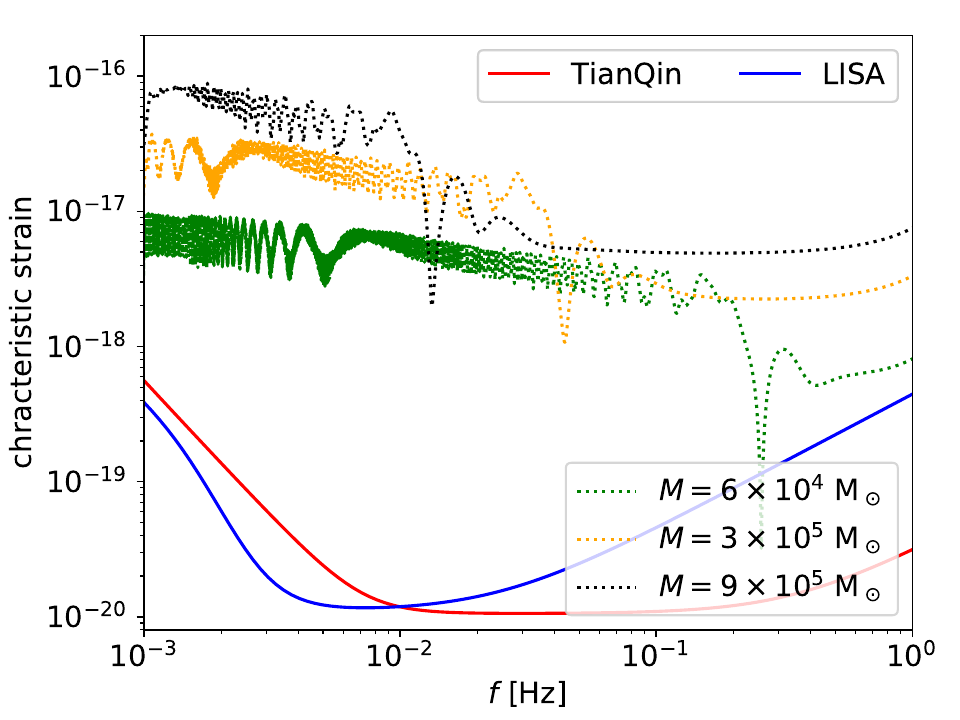}
    \caption{
        The characteristic strain for BBHs of masses $M=6\times10^4\,{\rm M_\odot}$ (green), $M=3\times10^5\,{\rm M_\odot}$ (orange), and $M=9\times10^5\,{\rm M_\odot}$ (black) over the noise curves of TianQin (red) and LISA (blue). The signals are all cut at the time of the merger.
    }\label{fig:stra}
\end{figure}

\fref{fig:stra} shows the characteristic strain $h_c(f) := 2f|\tilde{h}(f)|$ of the inspiral and merger phase for BBHs of the masses $6\times10^4\,{\rm M_\odot}$, $3\times10^5\,{\rm M_\odot}$, and $9\times10^5\,{\rm M_\odot}$ at a luminosity distance of $100\,{\rm Mpc}$ and a sky localization $(\text{RA},\text{dec})=(13^\text{h},+30^\circ)$, corresponding roughly to the position of the Coma Cluster, plotted over the noise curves of TianQin and LISA~\cite{torres-orjuela_huang_2023}. We adopt the antenna pattern functions introduced in Wang et al. (2019)~\cite{wang_jiang_2019} and Klein \& Barausse (2016)~\cite{klein_barausse_2016} for TianQin and LISA, respectively, assuming the polarization angle to be zero but ignore any intrinsic data gaps for both detectors~\cite{tianqin_2016,lisa_2017,lisa_2022c}. The last assumption is justified by the fact that the duration of the signals in the band we consider is shorter than the on-times of TianQin. Moreover, we consider only frequencies above $1\,{\rm mHz}$ because \texttt{NRHybSur3dq8} does not allow lower frequencies for sources of these masses. Considering lower frequencies, however, would only further improve the results obtained and hence we are considering a pessimistic case.

We see in \fref{fig:stra} that the inspiral and merger phases of these sources lie inside the bands of TianQin and LISA while their strain is also well above the noise of the detectors. The SNR of these sources is calculated as introduced in \sref{sec:det} where we find that in TianQin the SNR is around 4300, 14100, and 26500 while in LISA the SNR is around 16300, 59000, and 129800 for $M=6\times10^4\,{\rm M_\odot}$, $M=3\times10^5\,{\rm M_\odot}$, and $M=9\times10^5\,{\rm M_\odot}$, respectively. We see that BBHs with the masses considered at a luminosity distance of $100\,{\rm Mpc}$ are going to be detected with very high SNRs by TianQin and LISA. Therefore, we could, in principle, consider sources at much longer distances and still obtain a strong signal for the inspiral and merger. However, as we will see later the limiting factor for the distance we consider is the ability to detect the QNMs accurately enough.

AI detectors require some time to be switched into the resonant mode and also detailed information about the frequency that is going to be detected. Therefore, it is necessary to detect the target source and perform the data analysis some time in advance. We start our ringdown analysis for times $t>20MG/c^3$ where QNMs can be treated as linear solutions of a perturbed BH~\cite{london_2018}. These times correspond to around $6\,{\rm s}$, $30\,{\rm s}$, and $89\,{\rm s}$ after the merger for total masses of $6\times10^4\,{\rm M_\odot}$, $3\times10^5\,{\rm M_\odot}$, and $9\times10^5\,{\rm M_\odot}$, respectively. As we discussed in our previous paper~\cite{torres-orjuela_2023}, the time to switch the detector into resonant mode should be at least $30\,{\rm s}$ but longer time gaps are also adequate. Therefore, in the most ideal case where a time of $30\,{\rm s}$ to switch the detector is assumed and the full inspiral-merger phase shall be detected, it is necessary to consider BBHs with a total mass of at least $3\times10^5\,{\rm M_\odot}$. However, in the lowest mass case where $M=6\times10^4\,{\rm M_\odot}$ the source spends over 50 days in the band, and even for the highest mass case where $M=9\times10^5\,{\rm M_\odot}$ the source is more than half a day in the band. Considering that in the case discussed, not only analyzing the data and switching the detector is required but also communication between different detectors, we assume a more conservative time gap than before and set the time between the end of the detection of the inspiral-merger phase and the start of the detection of the QNMs to be $60\,{\rm min}$.

\subsection{Parameter estimation from space-based laser interferometer detectors}\label{sec:para}

In our analysis, we assume the inspiral and merger phases of the BBH are detected by TianQin or LISA which are then used to estimate the parameters of the source relevant for the detection of the QNMs. Therefore, we study in this section how accurately the total mass of the binary in the observer frame $M$, its mass ratio $q$, and the spin magnitude of the primary BH $\hat{a}_1$ can be detected. Note that we only consider the spin of the primary BH because its impact on the waveform and the remnant BH is more relevant and thus the results obtained represent the upper limit for both spins. Moreover, we only consider its magnitude because we restrict our analysis to non-precessing binaries. As we assume a time gap of $60\,{\rm min}$ between the end of the detection in TianQin or LISA and the start of the ringdown after $t>20MG/c^3$, we subtract the last $60\,{\rm min}-20MG/c^3$ of the data in TianQin/LISA. We consider a BBH with an average inclination of $60^\circ$ at a luminosity distance of $100\,{\rm Mpc}$ and a sky localization $(\text{RA},\text{dec})=(13^\text{h},+30^\circ)$, while setting the spin magnitude of the secondary black hole to $0.5$.

\begin{figure*}
\includegraphics[width=0.98\textwidth]{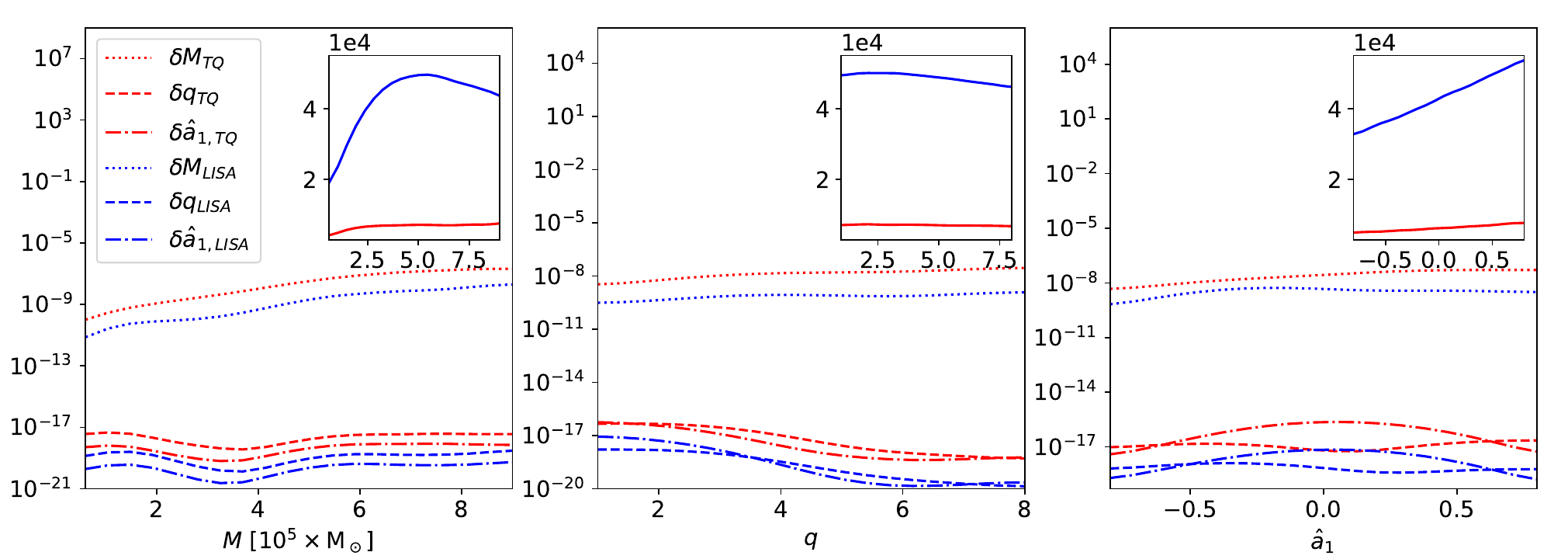}
    \caption{
        The relative error for the total mass $M$ (dotted lines), the mass ratio $q$ (dashed lines), and the spin magnitude of the primary BH $\hat{a}_1$ (dashed-dotted lines) estimated from the inspiral and merger phases of the BBH in TianQin (red) and LISA (blue). The left plot shows how the relative errors vary as a function of $M$, the center plot as a function of $q$, and the right plot as a function of $\hat{a}_1$. The SNR as a function of the same parameters is shown in the inset plots.
    }\label{fig:errim}
\end{figure*}

We estimate the relative errors for the total mass, the mass ratio, and the spin magnitude of the primary BH assuming the BBH is detected by TianQin or LISA using a Fisher matrix analysis. In \fref{fig:errim}, we show these relative errors as functions of the same parameters. We vary only one parameter at the same time and fix the two other parameters to the fiducial values $M=5\times10^5\,{\rm M_\odot}$, $q=4$, or $\hat{a}_1=0.5$. We see that for low masses $M$ is constrained down to an order of $10^{-10}$ and $10^{-12}$ for TianQin and LISA, respectively. For high masses, the total mass is constrained to an order of $10^{-7}$ for TianQin and an order of $10^{-8}$ for LISA. The variation of the detection accuracy of $M$ for a different mass ratio is only of around one order of magnitude reaching its maximum of around $10^{-8}$ in TianQin for $q\gtrsim3$ and of around $10^{-10}$ in LISA for $3\lesssim q\lesssim5$. $\delta M$ as a function of $\hat{a}_1$ also varies by only roughly one order of magnitude having a maximum of around $10^{-8}$ in TianQin for positive $\hat{a}_1$ while its maximum in LISA is of around $10^{-9}$ for $\hat{a}_1\approx-0.3$. The detection accuracy for $q$ and $\hat{a}_1$ as functions of $M$ vary less than the accuracy of the total mass although they have quite pronounced minima around $10^3-10^4\,{\rm M_\odot}$. $\delta q$ and $\delta\hat{a}_1$ tend to have their highest values for low masses $M\lesssim10^2\,{\rm M_\odot}$ and high masses $M\gtrsim10^6\,{\rm M_\odot}$ being of around $10^{-18}$ for TianQin and $10^{-19}$ for LISA, and of $10^{-18}$ for TianQin and $10^{-20}$ for LISA, respectively. The mass ratio is constrained the least for equal mass binaries being of the order $10^{-16}$ for TianQin and two orders of magnitude less for LISA as well as for high absolute values of $\hat{a}_1$ reaching around $10^{-17}$ for TianQin and $10^{-19}$ for LISA. $\delta\hat{a}_1$ is constrained the least for small $|\hat{a}_1|$ being of the order $10^{-16}$ and $10^{-17}$ for TianQin and LISA, respectively. The constraint on the spin magnitude of the primary BH is also the weakest for small mass ratios going up to $10^{-17}$ for TianQin and one order of magnitude below for LISA.

The better constraints by LISA are mainly due to LISA having a higher SNR for massive BBH which is a well-known fact~\cite{torres-orjuela_huang_2023}. Nevertheless, the total mass is constrained with lower accuracy for higher masses despite having a higher SNR because a higher mass also leads to the source having a shorter inspiral time in the band which is the phase that constrains the total mass the most. Moreover, we see that for LISA the SNR even decreases again for the highest masses which can be attributed to the relatively long gap of $60\,{\rm min}$ that leads to omitting a significant part of the merger phase -- in particular, for high-mass sources. However, if we could simulate the inspiral from when it enters the band and not only from $1\,{\rm mHz}$, the difference between the accuracy for low-mass sources and high-mass sources would be reduced. The improved accuracy for $q$ and $\hat{a}_1$ as the mass ratio increases despite the SNR going down -- which is more evident for LISA but also the case for TianQin -- can be understood from the fact the sources with high $q$ tend to be longer in the band which allows tracking the effects of the different parameters better. Understanding the dependence of the detection accuracy from $\hat{a}_1$ is more subtle as it does not only depend on the spin magnitude but also the orientation of the spin relative to the angular momentum of the source and the spin of the secondary BH. However, we see that the accuracies of $\hat{a}_1$ and $M$ tend to be better when $\hat{a}_1$ has a small magnitude while having an opposite trend for the detection accuracy of $q$.

We point out that the detection accuracy will improve if the time gap can be reduced. Moreover, detection could be further improved when considering joint detection by TianQin and LISA~\cite{torres-orjuela_huang_2023}. Nevertheless, we use the results from single detection with the $60\,{\rm min}$ gap to be conservative. It should be noted that the detection errors estimated here are orders of magnitude smaller than the numerical errors from \texttt{NRSur3dq8Remnant} which we use to obtain the parameters of the remnant. However, it can be expected that the numerical errors from \texttt{NRSur3dq8Remnant} and similar remnant models will greatly decrease in the coming years; hence we focus on the detection errors. As we discussed before, the detection errors can vary depending on the exact parameters of the BBH. For the rest of the paper, however, we approximate the detection errors using the maximal values found in our analysis independent of the actual configuration of the binary and set $\delta M = 10^{-7}$ and $\delta q = \delta\hat{a}_1 = 10^{-16}$. This way we obtain conservative estimates for the detection of QNMs with AI detectors.

Before estimating the accuracy with which the frequency of the QNMs can be predicted, we check if the time when the ringdown starts $t_{\rm RD}$ can be estimated accurately enough. We estimate the error in $t_{\rm RD}$ by adding the error from the time to coalescence $t_{\rm co}$ and the error from the time until the ringdown starts $t_0=20MG/c^3$
\begin{equation}
    \Delta t_{\rm RD} = \Delta t_{\rm co} + \Delta t_0.
\end{equation}
To evaluate how the error in the parameters impacts $\Delta t_{\rm co}$, we compare the time to coalescence we get when having the original parameters and when varying one of the parameters by the errors $\Delta M = M\delta M$, $\Delta q = q\delta q$, and $\Delta\hat{a}_1 = \hat{a}_1\delta\hat{a}_1$ while assuming the fiducial values $M=5\times10^5\,{\rm M_\odot}$, $q=4$, and $\hat{a}_1=0.5$. We obtain $t_{\rm co}$ from the time given by \texttt{NRHybSur3dq8} where we measure the time from the start of the gap. The error for the starting time of the ringdown is calculated assuming simple error propagation from the error of the mass $\Delta t_0 = t_0\delta M$.

\begin{figure}
\includegraphics[width=0.48\textwidth]{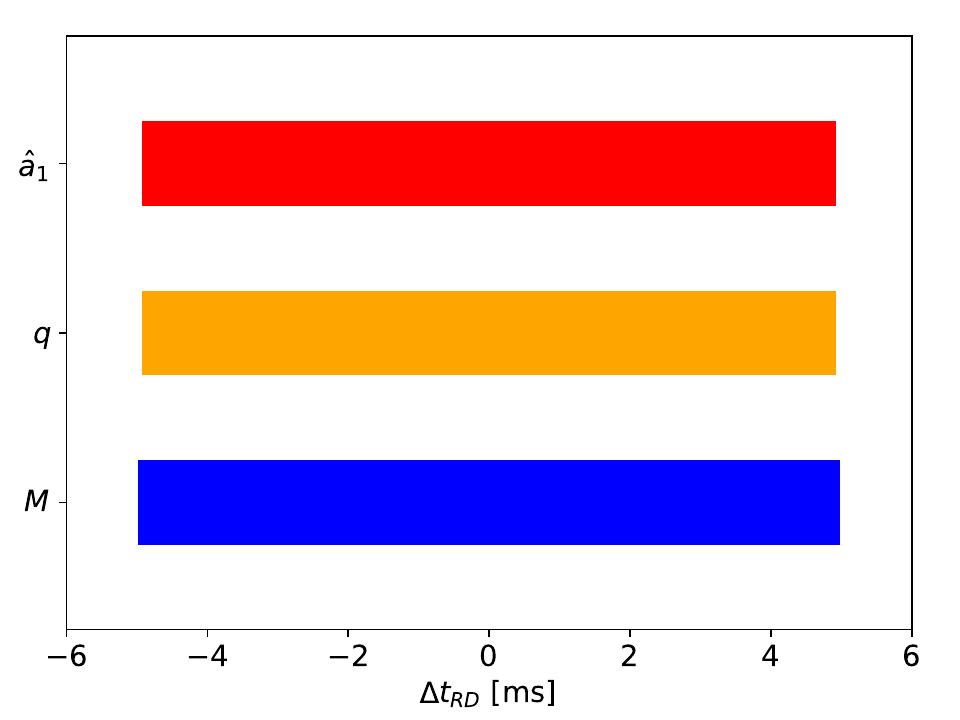}
    \caption{
        The error in the time when the ringdown starts $\Delta t_{\rm RD}$ resulting from the errors in the estimation of the total mass $\Delta M$, the mass ratio $\Delta q$, and the spin magnitude of the primary BH $\Delta\hat{a}_1$. Note that $\Delta t_{\rm RD}$ always depends on the error in the total mass as the starting time of the ringdown $t_0$ is a function of the total mass. See the text for more details.
    }\label{fig:timerr}
\end{figure}

\fref{fig:timerr} shows the error in the time when the ringdown starts $\Delta t_{\rm RD}$ for the different errors in $M$, $q$, and $\hat{a}_1$. As expected, $\Delta t_{\rm RD}$ has the strongest dependence on the total mass with an error of $\pm4.98\,{\rm ms}$. For $q$ and $\hat{a}_1$ we have $\Delta t_{\rm RD}\approx\pm4.92\,{\rm ms}$. The $\Delta t_{\rm RD}$ for the mass ratio and the spin is the same as it is completely dominated by the error in $t_0$. As a matter of fact, the $\Delta t_{\rm co}$ induced by $\Delta q$ and $\Delta\hat{a}_1$ are zero as the actual error is smaller than the time steps used in the waveform model. The error in the time when the ringdown starts is six orders of magnitude smaller than the time of the gap which is $60\,{\rm min}$, and hence does not affect the analysis of the ringdown. Even for a much shorter time gap of the order of seconds $\Delta t_{\rm RD}$ would be no significant hindrance for the analysis proposed, in particular, because a shorter time gap would result in a longer detection of the inspiral-merger phase which would further reduce the parameter estimation errors.

\begin{figure}
\includegraphics[width=0.48\textwidth]{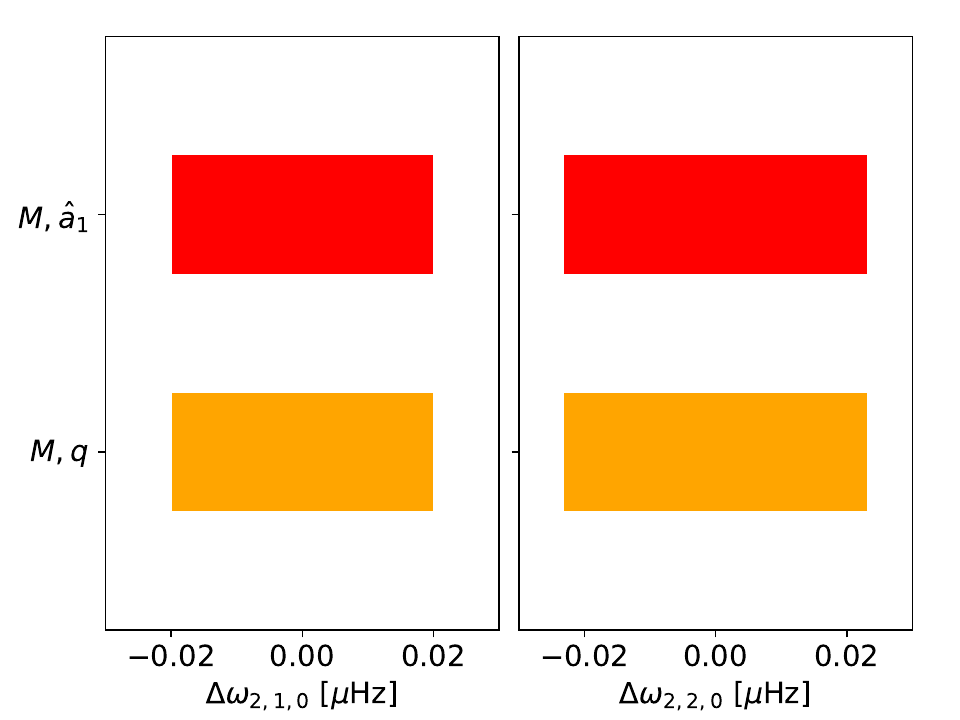}
    \caption{
        The error in the frequencies of the QNMs resulting from the errors in the estimation of the total mass $\Delta M$, the mass ratio $\Delta q$, and the spin magnitude of the primary BH $\Delta\hat{a}_1$. The left plot shows the error for the mode $(2,1,0)$ while the right plot shows the error for the mode $(2,2,0)$. The first line shows the combined error resulting from $\Delta M$ and $\Delta\hat{a}_1$ while the second line shows the combined error resulting from $\Delta M$ and $\Delta q$
    }\label{fig:frerr}
\end{figure}

As the last step in this subsection, we estimate the accuracy with which the frequencies of the QNMs $(2,1,0)$ and $(2,2,0)$ can be predicted from the parameter estimation using TianQin or LISA detections of the inspiral-merger phase. To estimate the error induced by $\Delta q$ and $\Delta\hat{a}_1$ we run the remnant model \texttt{NRSur3dq8Remnant} using the fiducial values as well as the values containing the error. The error from $\Delta M$ is estimated using error propagation in the analytical expressions of the QNMs. \fref{fig:frerr} shows the estimation error for the frequency of the $(2,1,0)$-mode and the $(2,2,0)$-mode. The error for the total mass and the two other parameters are shown combined as $\Delta M$ always influences the error in the frequency. As a matter of fact, $\Delta M$ dominates the error in the frequency. We find that the error for the QNM $(2,1,0)$ is around $0.020\,{\rm \mu Hz}$ while the error for the QNM $(2,2,0)$ is around $0.023\,{\rm \mu Hz}$. The error is symmetric as it does not significantly depend on whether the parameters are overestimated or underestimated. Moreover, note that the higher total error of $\omega_{2,2,0}$ can be attributed to it being bigger as the relative error of the two frequencies is essentially the same.

Independent of the specific mode, the error in the frequency is of the order $10^{-2}\,{\rm \mu Hz}$ and thus seven orders of magnitude smaller than the actual frequency. Note that since we use a Fisher matrix analysis to estimate the detection error, we are considering the error to be at $1-\sigma$ level. However, the error in the frequency of the QNMs as well as in the time when the ringdown starts is so small compared to the actual quantities that even when considering the $3-\sigma$ or $5-\sigma$ level, the error is still negligible. Therefore, these errors will not significantly limit the detection of the QNMs using the resonant mode of AI observatories.

\subsection{Black hole spectroscopy with atom interferometers}\label{sec:spect}

One of our goals is to estimate how accurately the QNMs of a merging BBH can be measured using the resonant mode of an AI observatory that is tuned to detect the two strongest QNMs $(2,2,0)$ and $(2,1,0)$. Before we proceed to this step, however, we analyze the detection of the late ringdown phase using the regular broadband mode of AION~\cite{aion_2020}. The benefits of starting with this analysis are twofold. First, we get a general idea of how well AI detectors in the ${\rm dHz}$-band can detect the QNMs of merging BBHs. Second, these results allow us to see if the detection of the QNMs can be improved when using the resonant mode. We point out that the analysis performed in this paper based on Berti et al. (2006)~\cite{berti_cardoso_2008} is rather simple and thus only approximative. More sophisticated approaches~\cite{ligo_virgo_2021e,pitte_baghi_2023,wang_shao_2023,wang_shao_2024} can and should be applied in real detections but we focus on this simpler formalism as we mainly want to highlight the possible gains from the detection of the QNMs using the resonant mode of AI detectors. 

Throughout this subsection, we use the same fiducial values as before: a BBH with an average inclination of $60^\circ$ at a luminosity distance of $100\,{\rm Mpc}$ and a sky localization $(\text{RA},\text{dec})=(13^\text{h},+30^\circ)$, while setting the spin magnitude of the secondary black hole to $0.5$. Moreover, we vary the total mass of the binary $M$, the mass ratio $q$, and the spin magnitude of the primary BH $\hat{a}_1$. However, we only vary one of these parameters at the same time and fix the two others to the values $M=5\times10^5\,{\rm M_\odot}$, $q=4$, or $\hat{a}_1=0.5$. Note that as we are considering non-precessing binaries, the spin of the remnant BH points in the same direction as the angular momentum of the binary, and hence the polarization angle remains zero after the merger. As AION is planned to be based in the UK, we set its location to be at $(51^\circ\,{\rm N},0^\circ\,{\rm W})$ (close to London) assuming the time of detection to be the March Equinox. We, further, use that AION is planned to be a vertical detector and ignore the rotation of the earth during the detection as the duration of the ringdown is only several seconds.

\begin{figure*}
\includegraphics[width=0.98\textwidth]{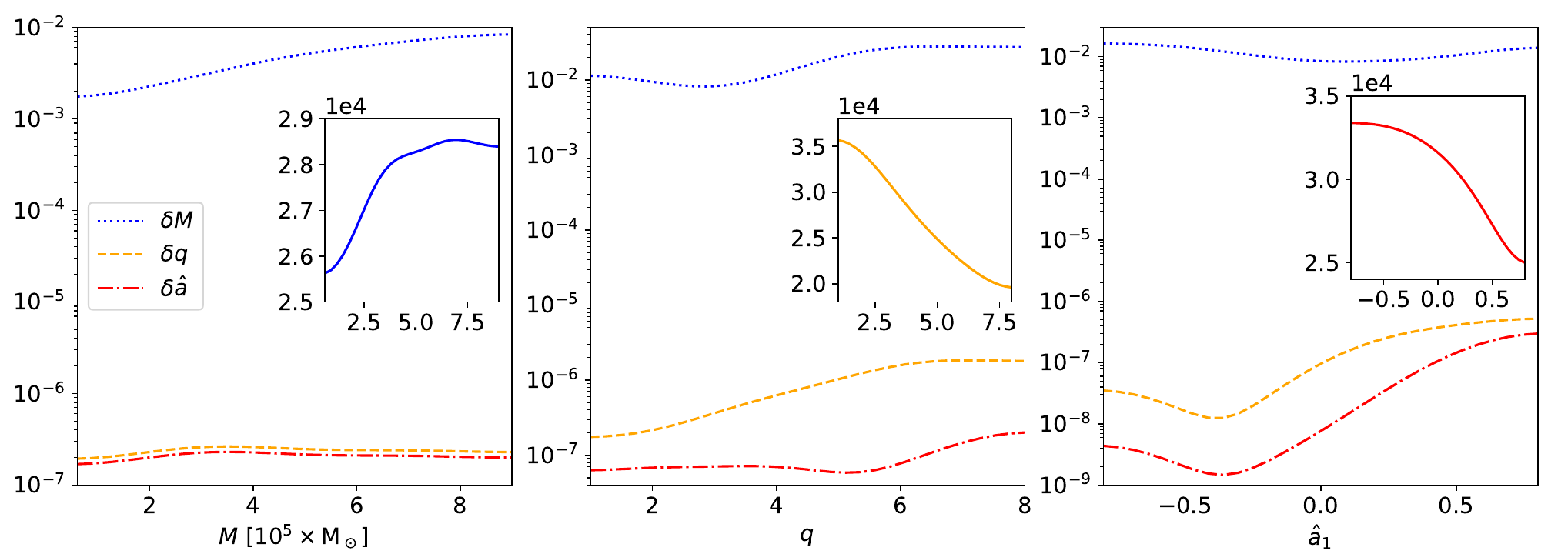}
    \caption{
        The relative error for the total mass $M$ (blue dotted line), the mass ratio $q$ (orange dashed line), and the spin magnitude of the primary BH $\hat{a}_1$ (red dashed-dotted line) estimated from the two dominant QNMs $(2,2,0)$ and $(2,1,0)$ using AION's broadband mode. The left plot shows how the relative errors vary as a function of $M$, the center plot as a function of $q$, and the right plot as a function of $\hat{a}_1$. The inset plots show the SNR as a function of the same parameters.
    }\label{fig:errbb}
\end{figure*}

We start with an analysis of how accurately the parameters of the BBH can be constrained from the detection of the QNMs. We focus again on $M$, $q$, and $\hat{a}_1$ to compare these results with the constraints we get from the inspiral-merger phase which can be used to test General Relativity. Usually, when testing General Relativity it is more common to compare the predicted properties of the remnant BH to the ones measured from the QNMs but as we use the properties of the BBH to predict the remnant BH and calculate the QNMs such a test would be meaningless in our case. From \fref{fig:errbb}, we see that the total mass of the BBH can be constrained to a relative error between $10^{-3}$ and almost $10^{-2}$ for low mass sources and high mass sources, respectively. The mass can be constrained less accurately for higher $M$ despite these sources having a higher SNR because their QNMs have higher frequencies and longer damping times causing part of the signal to be out of the band. Furthermore, $\delta M$ is constrained to an order of $10^{-2}$ for $q$ and $\hat{a}_1$, having slightly better accuracy for low mass ratio sources ($q\lesssim5$) as well as for a small spin of the primary BH ($|\hat{a}_1|\lesssim0.2$). $\delta q$ and $\delta\hat{a}_1$ are both constrained to an order of $10^{-7}$ with only small variation for different source masses. For a varying mass ratio, $\delta q$ is constrained to an order of $10^{-7}$ if $q\lesssim5$ and to an older of $10^{-6}$ for higher mass ratios because higher mass ratio sources have a lower SNR than sources with a low mass ratio. $delta\hat{a}_1$ as a function of $q$ is constrained to an order of $10^{-8}$ if $q\lesssim6$ and goes up to an order of $10^{-7}$ for higher mass ratios where the difference can be attributed again to the dependence of the SNR on $q$. $\delta q$ and $\delta\hat{a}_1$ show similar behavior as functions of $\hat{a}_1$ having a minimum at $\hat{a}_1\approx-0.35$ and a maximum for high spins above $0.5$, where the minimum can be attributed to the QNMs having similar frequencies for this value of $\hat{a}_1$ (cf. \fref{fig:fre}) and the maxima anticorrelate with the SNR going down for higher spin values. $\delta\hat{a}_1$, however, is around $10^{-9}$ at its minimum and $\delta q$ is of the order $10^{-8}$ while both go up to an order of $10^{-7}$ at their maximum.

In general, we see that the constraints from the ringdown phase are at least five orders of magnitude worse than the worst constraints from the inspiral-merger phase for the total mass,  ten orders of magnitude worse for the mass ratio, and seven orders of magnitude worse for the spin of the primary BH. Therefore, we see that the accuracy of tests comparing the constraints from the inspiral-merger phase to the ringdown phase is mainly limited by the accuracy we get from the latter one. Nevertheless, we would like to point out that these accuracies are still orders of magnitude better than the current results obtained by the LIGO-Virgo-KAGRA Collaboration~\cite{ligo_virgo_2021e} and comparable or better than the results expected with future space-based laser interferometer detectors or similar multi-band tests~\cite{shi_bao_2019,pitte_baghi_2023,carson_yagi_2020a,carson_yagi_2020b,jani_shoemaker_2020}.

\begin{figure*}
\includegraphics[width=0.98\textwidth]{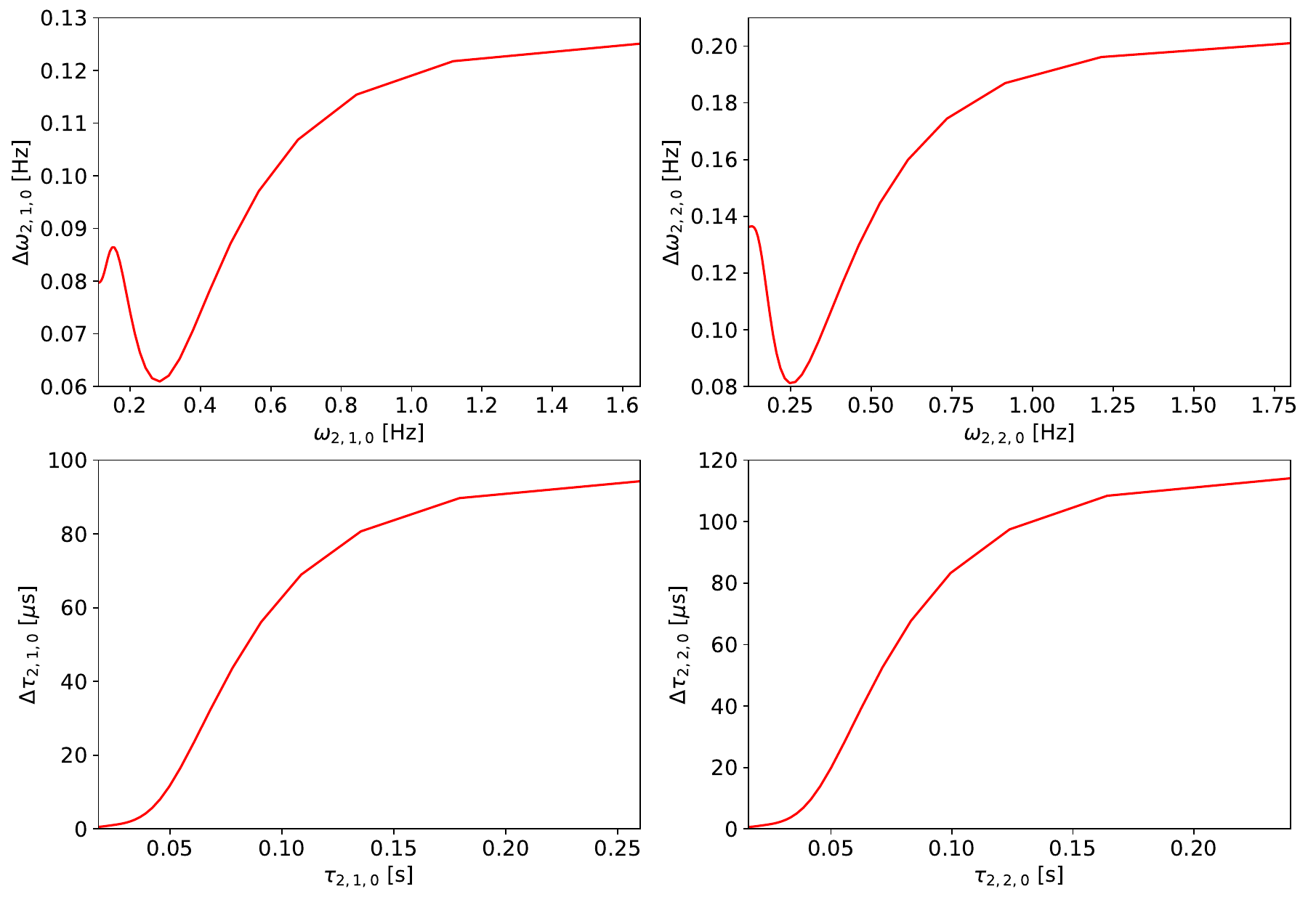}
    \caption{
        The absolute error in the frequency $\Delta\omega_{\ell,m,n}$ (top) and the damping time $\Delta\tau_{\ell,m,n}$ (bottom) of the $(2,1,0)$-mode (left) and the $(2,2,0)$-mode (right) using AION's broadband mode.
    }\label{fig:errqb}
\end{figure*}

Although it is possible to construct a waveform model that depends on the parameters of the BBH such as the total mass, the mass ratio, and the spin magnitude of the primary BH -- as we did before -- these are not the quantities that are detected during the ringdown~\cite{berti_cardoso_2008,ligo_virgo_2021e}. Instead, we measure the frequency $\omega_{\ell,m,n}$ and the damping time $\tau_{\ell,m,n}$ of the QNMs. We test how well they can be detected by treating the waveform model as a phenomenological model with the frequencies and damping times as its parameters. \fref{fig:errqb} shows how accurately $\omega_{2,2,0}$, $\omega_{2,1,0}$, $\tau_{2,2,0}$, and $\tau_{2,1,0}$ are detected. The parameters of the BBH are the fiducial values used in the previous analysis where we vary the total mass of the binary to obtain the different frequencies and damping times. Therefore, the SNR is the same as shown in the inset plot of the left subplot in \fref{fig:errbb} but lower frequencies/shorter damping times correspond to higher total masses while higher frequencies/longer damping times correspond to lower total masses.

\fref{fig:errqb} suggests that lower frequencies below roughly $0.75\,{\rm Hz}$ and shorter damping times below around $0.1\,{\rm s}$ are detected significantly better then bigger values. However, due to the corresponding short damping times, these results are less reliable and should be taken cautiously. Nevertheless, for frequencies above $0.75\,{\rm Hz}$ we see that $\omega_{2,1,0}$ is detected with a $1-\sigma$ error between roughly $0.1\,{\rm Hz}$ and $0.13\,{\rm Hz}$ while $\omega_{2,2,0}$ is constrained with an accuracy of roughly $0.18-0.2\,{\rm Hz}$. Therefore, the frequencies of the two dominant modes are constrained to at least a $3-\sigma$ level for low frequencies $\omega_{\ell,m,n,}\approx0.75\,{\rm Hz}$ and better than the $5-\sigma$ level for high frequencies $\omega_{\ell,m,n,}\gtrsim1\,{\rm Hz}$. We further see that damping times above $0.1\,{\rm s}$ are detected with an accuracy of around $50-95\,{\rm \mu s}$ and around $80-115\,{\rm \mu s}$ at the $1-\sigma$ level for $\tau_{2,1,0}$ and $\tau_{2,2,0}$, respectively. Therefore, the damping times are well constrained at the $5-\sigma$ level for all values.

Next, we proceed to study the detection of the QNMs using the resonant mode of AI detectors. As we focus on the two dominant modes $(2,1,0)$ and $(2,2,0)$, we define $f_r = (\omega_{2,1,0}+\omega_{2,2,0})/2$ and $Q = \lfloor f_r/|\omega_{2,1,0}-\omega_{2,2,0}|\rfloor$; cf. \sref{sec:det} for their definition. Note that we mix here the symbols $f$ and $\omega$ for the frequency to stick to the conventions of the different fields but they refer to the same quantities. Moreover, we point out that cutting out the frequencies outside of the band in the resonant mode leads to numerical instabilities during the analysis. Therefore, we also consider frequencies outside the band of the resonant mode although their contribution to the signal becomes negligible as the sensitivity of the detectors there is multiple orders of magnitude worse than inside the band.

\begin{figure*}
\includegraphics[width=0.98\textwidth]{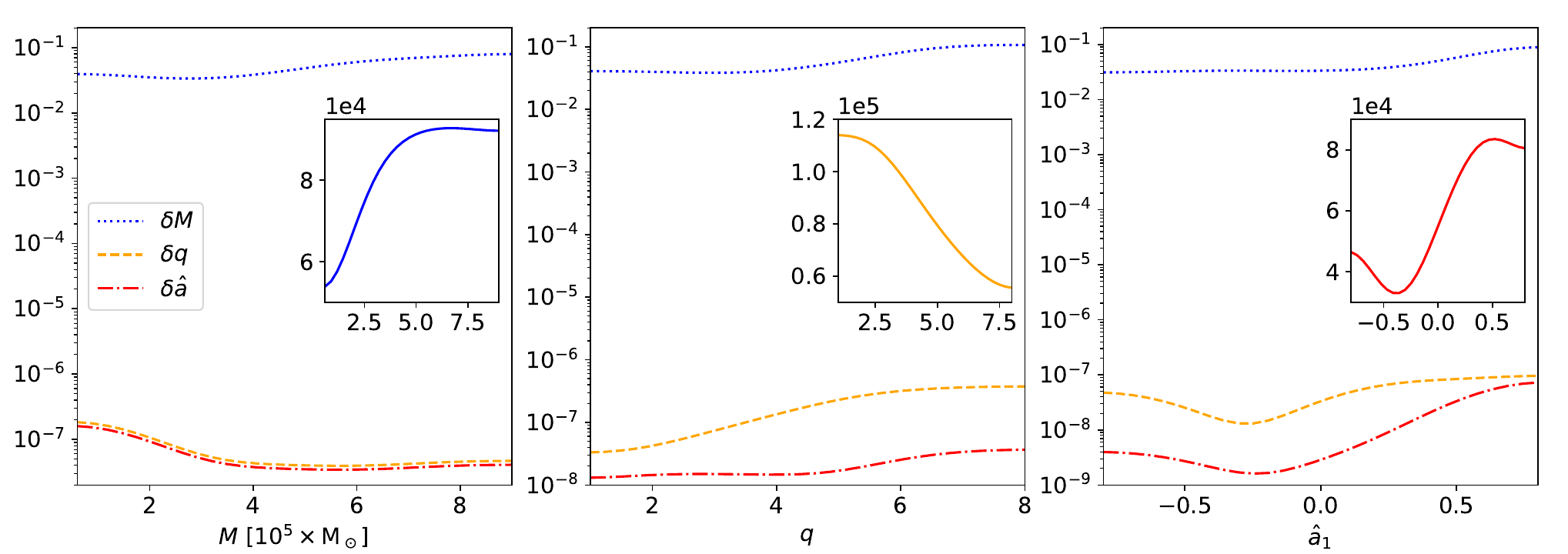}
    \caption{
        The relative error for the total mass $M$ (blue dotted line), the mass ratio $q$ (orange dashed line), and the spin magnitude of the primary BH $\hat{a}_1$ (red dashed-dotted line) estimated from the two dominant QNMs $(2,2,0)$ and $(2,1,0)$ using AION's resonant mode. The left plot shows how the relative errors vary as a function of $M$, the center plot as a function of $q$, and the right plot as a function of $\hat{a}_1$. The inset plots show that SNR as a function of the same parameters.
    }\label{fig:errrs}
\end{figure*}

We show in \fref{fig:errrs} the relative errors for the parameters of the BBH $M$, $q$, and $\hat{a}_1$ when using the resonant mode for detection. We see that the relative error in the total mass $M$ as a function of the three parameters shows similar behavior to the detection using the broadband mode but is now always constrained at a level of $10^{-2}$. $\delta q$ and $\delta\hat{a}_1$ as functions of $M$ are constrained to an order of $10^{-7}$ for low mass sources $M\lesssim10^5\,{\rm M_\odot}$ but to an order of $10^{-8}$ for higher mass sources which roughly anticorrelates with the SNR. The better constraints of the mass ratio and the spin of the primary BH compared to the detection with the broadband can be mainly attributed to having a higher SNR when using the resonant mode. The behavior of $\delta q$ and $\delta\hat{a}_1$ as functions of $q$ is similar to the case of the broadband mode but $\delta\hat{a}_1$ is always constrained to an order of $10^{-8}$ while $\delta q$ is constrained to an order between $10^{-8}$ and $10^{-7}$. The higher accuracy can once again be attributed to having a higher SNR in the resonant mode than in the broadband mode. The SNR as a function of $\hat{a}_1$ is for the resonant mode once again higher although its functional behavior shows significant differences. Nevertheless, $\delta q$ and $\delta\hat{a}_1$ as functions of $\hat{a}_1$ have again minima at $\hat{a}_1\approx-0.35$ and a maximum for high spins. While the minima are at around $10{-9}$ and $10^{-8}$ for $\delta\hat{a}_1$ and $\delta q$, respectively, their maxima are only at an order of $10^{-8}$.

We see that when using the resonant mode, the parameters of the BH can be constrained to the same or a better level as when using the broadband mode. The only exception is $\delta M$ as a function of the total mass which is constrained to almost one order of magnitude better in the broadband mode. The biggest gains are for $\delta q$ and $\delta\hat{a}_1$ which are constrained better by almost one order of magnitude for all values of $q$ considered as well as four sources with a high primary spin $\hat{a}_1\gtrsim0.5$. As we will explain later in more detail, the difference in the detection of $M$, $q$, and $\hat{a}_1$ depends on how they impact the frequencies and damping times of the QNMs as using the resonant mode improves the detection of the latter while diminishing the accuracy for the frequencies.

\begin{figure*}
\includegraphics[width=0.98\textwidth]{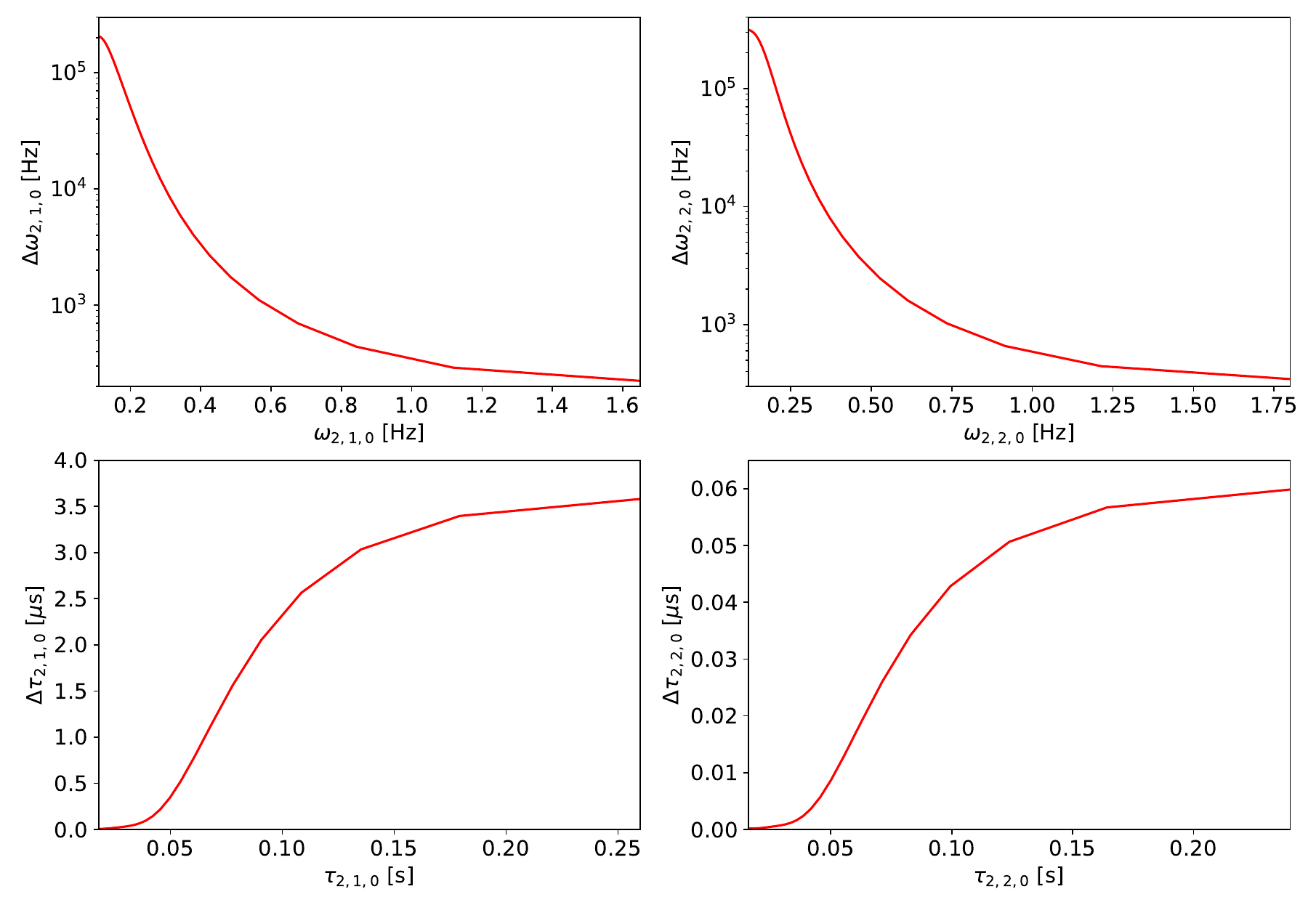}
    \caption{
        The absolute error in the frequency $\Delta\omega_{\ell,m,n}$ (top) and the damping time $\Delta\tau_{\ell,m,n}$ (bottom) of the $(2,1,0)$-mode (left) and the $(2,2,0)$-mode (right) using AION's resonant mode.
    }\label{fig:errqr}
\end{figure*}

From \fref{fig:errqr}, we see that using the resonant mode the frequency of the QNMs is basically not constrained having absolute errors of hundreds to hundred-thousands of ${\rm Hz}$ and thus two to six orders of magnitude bigger than the actual value. In contrast, the damping times are constrained very accurately when using the resonant mode. Focusing again on damping times above $0.1\,{\rm s}$, we see that $\tau_{2,1,0}$ is contained with a $1-\sigma$ error between $2.5\,{\rm \mu s}$ and $3.5\,{\rm \mu s}$ while the total error of $\tau_{2,2,0}$ is roughly $0.45-0.6\,{\rm \mu s}$. Therefore, using the resonant mode of AI detectors allows us to greatly improve the detection accuracy of the damping time of the QNMs -- by one order of magnitude for $\tau_{2,1,0}$ and by three to four orders of magnitude for $\tau_{2,2,0}$ -- but at the cost of not being able to constrain their frequency.

To understand why the frequency of the QNMs is constrained worse when using the resonant mode but the detection of the damping time improves, it is important to realize that QNMs have a constant frequency but are not monochromatic waves in a strict sense. That means they do not transform to $\delta$-functions in Fourier space but to functions of the form~\cite{berti_cardoso_2006}
\begin{equation}
    \frac{1/\tau_{\ell,m,n}}{(1/\tau_{\ell,m,n})^2+(\omega-\omega_{\ell,m,n})^2}
\end{equation}
when only considering positive frequencies. This function has a single maximum at $\omega_{\ell,m,n}$ while its slope is determined by $\tau_{\ell,m,n}$. Choosing the $Q$-factor so that the two frequencies $\omega_{2,1,0}$ and $\omega_{2,2,0}$ just fit in the band leads to one slope of each mode to be cut out. While detecting only one slope for each QNM with high accuracy allows an improved constraint on the damping time it makes it difficult to determine the maxima of the function and thus the frequencies.

\begin{figure*}
\includegraphics[width=0.98\textwidth]{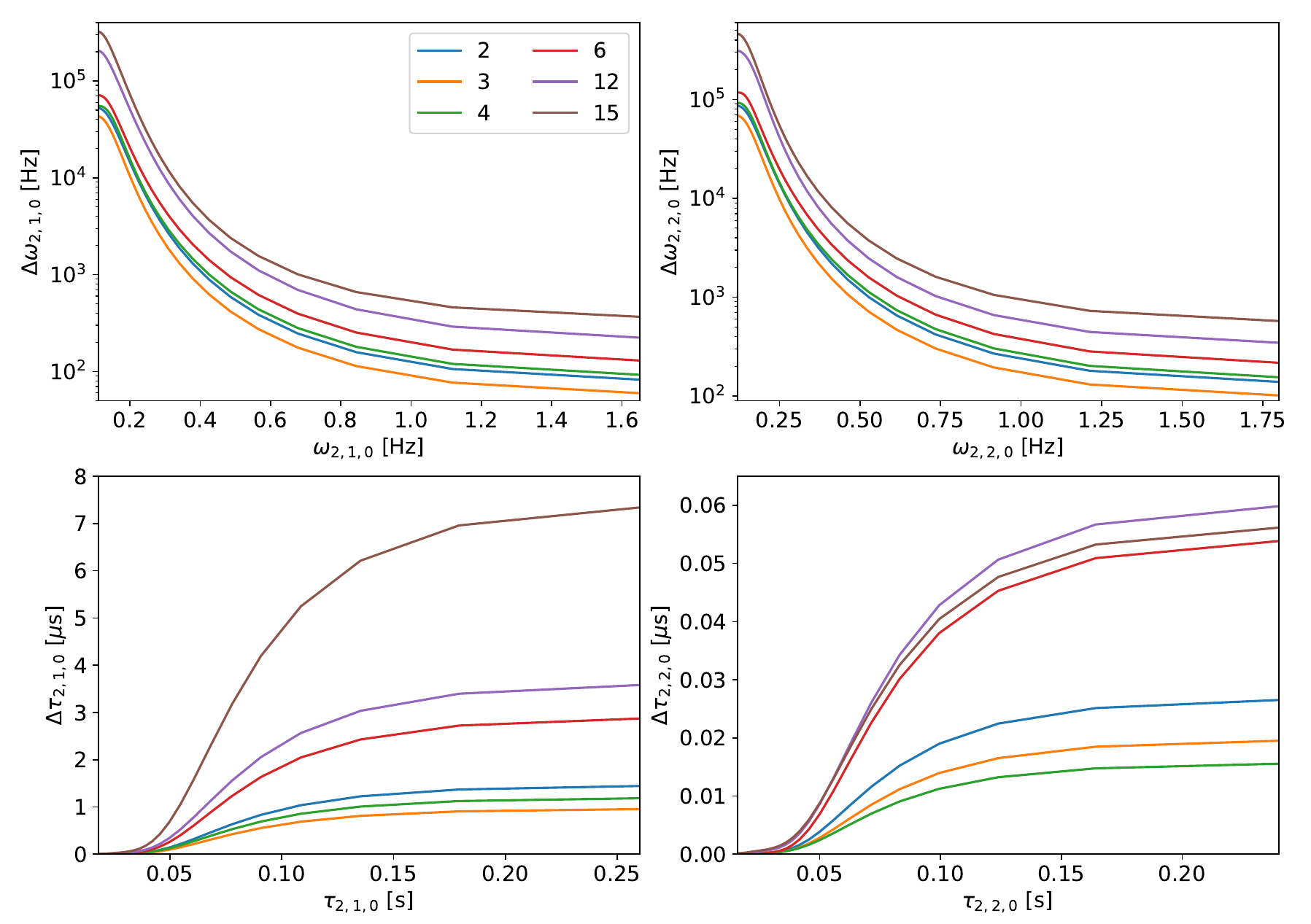}
    \caption{
        The absolute error in the frequency $\Delta\omega_{\ell,m,n}$ (top) and the damping time $\Delta\tau_{\ell,m,n}$ (bottom) of the $(2,1,0)$-mode (left) and the $(2,2,0)$-mode (right) using AION's resonant mode. Different colors represent different $Q$'s used in the resonant mode.
    }\label{fig:errqrd}
\end{figure*}

In our case, the value of the $Q$-factor depends strongly on the difference between $\omega_{2,1,0}$ and $\omega_{2,2,0}$. It can reach high values of up to 40 for small differences as in the case of small negative $\hat{a}_1$ (cf. \fref{fig:fre}) but in most cases it has a value of 12. As discussed in the previous paragraph, taking the highest possible value of $Q$ basically hinders measuring the frequency of the QNMs. Therefore, we explore next how well the frequency and the damping time of the QNMs can be constrained when using the resonant mode with different values $Q=2,3,4,6,12,15$. In \fref{fig:errqrd}, we see that going to a $Q$-factor bigger than 12 tends to make detection worse as a bigger chunk of the data is cut out. Going to smaller $Q$ tends to improve detection but taking, e.g., $Q=6$ only brings little improvement as doubling the size of the detection band does not make up for a reduction of the detection sensitivity by a factor of two. We see that the best detection accuracy is obtained for $Q$ either three or four. In the first case, $\tau_{2,1,0}$ is constrained the best reaching a detection accuracy below $1\,{\rm \mu s}$ and thus roughly three times better than for $Q=12$, while in the latter case $\Delta\tau_{2,2,0}$ is constrained to around $0.01\,{\rm \mu Hz}$ which is around five times better than for $Q=12$. Nevertheless, for none of the $Q$ shown the frequencies can be detected with accuracies similar to those obtained using the broadband mode. Therefore, it becomes clear that using the resonant mode can bring significant benefits when detecting the damping time of QNMs but at the cost of losing almost all information about their frequency.

\section{Conclusions}\label{sec:con}

In this paper, we study the detection of a merging BBH in the intermediate mass range $M=6\times10^4-9\times10^5\,{\rm M_\odot}$ and of its QNMs. We apply a Fisher matrix analysis to determine how accurately the parameters of the BBH can be estimated by the detection of the inspiral and merger phases in the ${\rm mHz}$-band with space-based laser interferometer detectors TianQin and LISA. Furthermore, we study the detection of the QNMs in the ${\rm dHz}$-band by the ground-based AI detector AION using a Fisher matrix analysis again. In agreement with previous results, we find that the detection of the inspiral-merger phase by TianQin and LISA will allow measuring the total mass, the mass ratio, and the spin magnitude of the primary BH with accuracies of at least $10^{-7}$, $10^{-16}$, and $10^{-16}$, respectively, for a BBH at the location of the Coma Cluster. We further explore how well the start time of the late ringdown (linear regime) and the frequency of the QNMs can be predicted from the detection of the inspiral-merger phase assuming that roughly the last $60\,{\rm min}$ of the merger are not detected to account for the time required to analyze the data and to switch an AI detector into resonant mode. We find that the starting time of the ringdown can be predicted with an accuracy of milliseconds while the frequency of the QNMs can be predicted to an order of micro-Hertz. Thus it is possible to predict the properties of the ringdown with high accuracy which will allow us to tune the resonant mode of AI detectors to the detection of specific QNMs if desired.

We study the detection of the QNMs by AI detectors using the regular broadband mode as well as the resonant mode. Assuming again the source to be in the intermediate mass range and at a distance of the Coma Cluster -- which is more or less the highest distance allowing accurate detection of the QNMs -- we explore how accurately the total mass, the mass ratio, and the spin magnitude of the primary BH can be determined from the ringdown phase. We find that the detection accuracy is five to ten orders of magnitude worse than the constraints from the inspiral-merger phase detected by space-based laser interferometer detectors. Nevertheless, the accuracy is still very good being of at least an order of $10^{-6}$ which will allow for accurate constraints on General relativity that far surpass current results by the LIGO-Virgo-KAGRA Collaboration. We also study the detection of the frequency of the QNMs and their damping time where we find that the first can be detected with an error below $0.125\,{\rm Hz}$ for the $(2,1,0)$-mode and below $0.2\,{\rm Hz}$ for the $(2,2,0)$-mode while the latter can be constrained to less than $95\,{\rm \mu s}$ and $115\,{\rm \mu s}$ for the QNMs $(2,1,0)$ and $(2,2,0)$, respectively. Therefore, using the broadband mode of AI detectors the frequency and the damping time of the dominant QNMs can be constrained well at the $3-\sigma$ to $5-\sigma$ level.

Furthermore, we explore how well the two dominant QNMs can be detected using the resonant mode of AI detectors. We find that the parameters of the BBH can be mostly determined at a similar level as when using the broadband mode having an improvement of almost one order of magnitude for the detection error of the mass ratio and the spin magnitude of the primary BH for all mass ratios considered and in the case of a high primary spin. Furthermore, we find that the detection error for the damping time of the QNMs can be greatly reduced -- by one order of magnitude for $\tau_{2,1,0}$ and by three to four orders of magnitude for $\tau_{2,2,0}$ -- when using the resonant mode but at the cost of not being able to constraint their frequency at all. The reason for the improved detection of the damping time with a loss of basically all information about the frequency is that using the resonant modes allows a higher detection sensitivity but in a narrower band. The improved detection sensitivity allows for determining the slope of the Fourier-transformed QNMs with higher accuracy resulting in better constraints of the damping time while the narrower band makes it difficult to determine the maxima of the Fourier-transforms which contain the information about the frequency.

Last, we study how changing the $Q$-factor of the resonant mode which is proportional to the gain in sensitivity and inversely proportional to the bandwidth impacts the detection error of the frequencies and the damping times of the QNMs. We find that $Q$-factors of three and four -- which are several times lower than $Q=12$, the highest possible factor that still allows the detection of the two dominant QNMs -- yield the highest improvement in the detection error of the damping time. Their detection improves by factors of three to five reaching accuracies of $1\,{\rm \mu Hz}$ for the QNM $(2,1,0)$ and of $0.01\,{\rm \mu Hz}$ for the $(2,2,0)$-mode. Although the detection of the frequencies also improves by similar factors they remain completely undetermined. Therefore, we conclude that using the resonant mode of AI detectors is beneficial when detecting the damping time of multiple QNMs but it comes at the cost of not being able to detect their frequency.

We point out that another option to use the resonant mode would be to focus on the detection of only one mode. Due to the very good accuracy with which the frequency of the QNMs can be predicted from the inspiral-merger phase, it would be relatively easy to center the detection band around the frequency of the mode considered and one would be free to set the $Q$-factor to an optimal value for detection. We did not explore this possibility in this paper, as tests of General Relativity often require the detection of at least two QNMs. Nevertheless, exploring this idea in the future is interesting, in particular, as it could allow the detection of subdominant modes that might not be detectable otherwise.

\begin{acknowledgments}
This work was partially supported by the Key Laboratory of TianQin Project (Sun Yat-sen
University), Ministry of Education (China).
\end{acknowledgments}

\section*{Conflict of interest}

The authors have no conflicts to disclose.

\section*{Data Availability Statement}

The data that support the findings of this study are available from the corresponding author upon reasonable request.

\bibliography{alebib}

\end{document}